\documentclass{article}

\usepackage{authblk}
\usepackage{graphicx}
\usepackage[font=small,labelfont=bf]{caption}

\usepackage[margin=2.5cm]{geometry}
\usepackage{hyperref}
\usepackage{amssymb}
\usepackage{amsmath}
\DeclareMathOperator*{\argmin}{arg\,min}
\usepackage{lineno}
\usepackage{floatrow}
\usepackage{multirow}
\usepackage{setspace}
\usepackage{caption}
\makeatletter
\def\@maketitle{\newpage
 \null
 \begin{flushleft}%
  {\LARGE \@title \par}%
  \vskip 1.5em
  {\lineskip .5em
   \@author}%
 \end{flushleft}%
 \par
 \vskip 1.5em}
 \makeatother

\renewenvironment{abstract}
 {\par\noindent\textbf{\abstractname.}\ \ignorespaces}
 {\par\medskip}

\providecommand{\keywords}[1]
{
  \small	
  \noindent Keywords: #1
}

\title{Quantifying the effect of X-ray scattering for data generation in real-time defect detection}
\author[a]{Vladyslav Andriiashen\thanks{
Corresponding author: Vladyslav Andriiashen, Computational Imaging, Centrum Wiskunde en Informatica, Amsterdam, 1098 XG, The Netherlands. E-mail: vladyslav.andriiashen@cwi.nl.
}}
\author[a,b]{Robert van Liere}
\author[a,c]{Tristan van Leeuwen}
\author[a,d]{Kees Joost Batenburg}

\affil[a]{Computational Imaging, Centrum Wiskunde en Informatica, Amsterdam, The Netherlands}
\affil[b]{Faculteit Wiskunde en Informatica, Technical University Eindhoven, Eindhoven, The Netherlands}
\affil[c]{Mathematical Institute, Utrecht University, Utrecht, The Netherlands}
\affil[d]{Leiden Institute of Advanced Computer Science, Leiden University, Leiden, The Netherlands}

\date{}
\begin{document}

%\linenumbers

\maketitle

\begin{abstract}
\\ \textbf{BACKGROUND:} X-ray imaging is widely used for the non-destructive detection of defects in industrial products on a conveyor belt. In-line detection requires highly accurate, robust, and fast algorithms. Deep Convolutional Neural Networks (DCNNs) satisfy these requirements when a large amount of labeled data is available. To overcome the challenge of collecting these data, different methods of X-ray image generation are considered.
\\ \textbf{OBJECTIVE:} Depending on the desired degree of similarity to real data, different physical effects should either be simulated or can be ignored. X-ray scattering is known to be computationally expensive to simulate, and this effect can greatly affect the accuracy of a generated X-ray image. We aim to quantitatively evaluate the effect of scattering on defect detection.
\\ \textbf{METHODS:} Monte-Carlo simulation is used to generate X-ray scattering distribution. DCNNs are trained on the data with and without scattering and applied to the same test datasets. Probability of Detection (POD) curves are computed to compare their performance, characterized by the size of the smallest detectable defect.
\\ \textbf{RESULTS:} We apply the methodology to a model problem of defect detection in cylinders. When trained on data without scattering, DCNNs reliably detect defects larger than 1.3~mm, and using data with scattering improves performance by less than 5\%. If the analysis is performed on the cases with large scattering-to-primary ratio ($1 < SPR < 5$), the difference in performance could reach 15\% (approx. 0.4~mm).
\\ \textbf{CONCLUSION:} Excluding the scattering signal from the training data has the largest effect on the smallest detectable defects, and the difference decreases for larger defects. The scattering-to-primary ratio has a significant effect on detection performance and the required accuracy of data generation.

\end{abstract}

\keywords{X-ray imaging, X-ray data generation, X-ray scattering, Deep learning, In-line inspection}

\section{Introduction}

\begin{figure}[ht]
%\begin{figure}[p]
\centering
\includegraphics[width=0.60\linewidth]{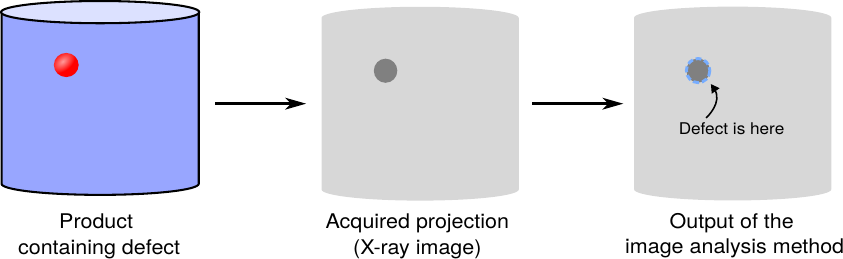}
\caption{General scheme of defect detection via X-ray imaging. An X-ray acquisition system is used to make a projection of the product of interest. Defects affect projection intensity even if they are inside the product, and can be detected by analyzing the projection.}
\label{defect_detection}
\end{figure}

X-ray imaging is used as a non-destructive testing technique suitable for objects of different shapes made of different materials \cite{ISO_xray, chen2021interfacial}. We focus on the task of defect detection: a product may contain an undesirable structure (an object made of different material or a void) that should be detected by an inspection system. 
Depending on the task, detection may only require indicating that a defect is present, or locating it and specifying its size and other properties.
Examples of defects in this context include spallation in metal alloys, blowholes in castings, razors in airplane baggage, bones in fish fillets, etc. Due to differences in density and chemical structure, defects affect an X-ray image (later referred to as a projection) of the product. Their presence can be detected by analyzing the projection with an algorithm or a human expert (Fig. \ref{defect_detection}). A projection can be acquired in tens of milliseconds, making X-ray imaging an in-line inspection technique. While individual projections can be inspected by a human expert, high throughput of data requires a fast and accurate defect detection algorithm. The main challenge in analyzing X-ray projections is the overlap between features of the object located at different depths. This complicated task is conventionally solved by human experts relying on application-specific knowledge.

In recent years, deep learning methods, such as Deep Convolutional Neural Networks (DCNNs), have been successfully applied in many areas of industrial product inspection \cite{mery2015computer, yang2020using, lee2021review}. Deep learning algorithms assume that the image processing task can be solved by a function with a large number of parameters. The values of these parameters are found by optimizing the result of the function for a large set of projections (referred to as training data). Compared to conventional image processing algorithms, DCNNs provide higher accuracy at the expense of interpretability. The high predictive power of DCNNs stems from their ability to generalize image patterns present in the training dataset. Such datasets, with a few exceptions \cite{mery2015gdxray}, are not available for X-ray defect detection. To overcome this barrier, experts from various fields have tried to simulate X-ray data using computational methods \cite{bellon2007artist, gong2018rapid}. Depending on the desired degree of similarity to real data, the methods range from ray-tracing to Monte-Carlo simulation of each individual X-ray photon.

The computational cost of the simulation algorithm for X-ray imaging comes from the models of X-ray interaction with objects and the detector. Most methods do not estimate the effect of X-ray photon scattering, since it significantly increases the computational complexity of the problem \cite{gong2019rapid}. Instead, only primary radiation - photons that pass through the object without scattering - is simulated. If the training data are generated using only the primary X-ray signal, the DCNN will miss the influence of scattered photons that are present in the real test data. Hence, a decrease in accuracy is possible. Due to the low interpretability of DCNNs, it is impossible to predict in advance how a change in the training data would affect performance on a wide range of test cases. Balancing the computational cost of the data generation approach and the resulting DCNN performance requires an accurate and robust metric of detection accuracy.

The Probability of Detection (POD) method \cite{georgiou2007pod} was introduced in non-destructive testing and evaluation to analyze the reliability of inspection techniques. Recently it has been applied to X-ray radiography and CT \cite{tyystjarvi2022automated, yosifov2022probability}. POD analysis originally addressed the problem of determining the smallest defect size that can be found with an inspection system under specified measurement conditions. The introduction of \textit{Probability} stems from the existence of many possible products with the same defect size, and only in a fraction of them the inspection is successful (e.g. due to the location of the defect). The POD curves aim to estimate these fractions by statistical analysis of the inspection results on the test dataset. The POD method can be applied to any inspection technique including deep learning methods. While the POD curves do not analyze the influence of the training set on the DCNN, they connect the performance of the networks to the application requirements.

We present a methodology that quantitatively evaluates the effect of X-ray scattering in the training data on the accuracy of DCNNs for defect detection in X-ray projections. To demonstrate its application, we perform computational experiments on a model problem consisting of the detection of a cavity inside a cylindrical object. Using a Monte-Carlo simulation algorithm, we generate two versions of the same training data that include and exclude scattered X-ray photons. These data are used to train two DCCNs. We use POD curves to evaluate the correlation between network accuracy and test projection properties. This allows us to separate the influence of training and test data properties and to highlight conditions under which simulation of scattering is crucial. We discuss how this methodology can be applied to other industrial problems and which task properties significantly influence the results. 

\section{Related work}

\subsection{Deep learning in real-time defect detection}

The problem of defect detection can take different forms depending on the desired output of the algorithm. \textit{Classification} returns a single label corresponding to the type of defect present in the projection (e.g., steel defects \cite{masci2012steel}). \textit{Object detection} returns bounding boxes for each defect and labels corresponding to their types (commonly used in baggage inspection \cite{aydin2018new} and quality control of metal details, such as aluminum castings \cite{parlak2023deep}). \textit{Segmentation} provides a set of pixel masks: each pixel of the projection is labeled if a defect is present there (e.g. spallation in aircraft engines \cite{bian2016multiscale}). There is a difference between \textit{semantic} (one mask for all defects of the same type) and \textit{instance} (different defects of the same type are separated) segmentation. This is not relevant for our model problem, where there is only a single defect.

In contrast to color photographs, there is a lack of publicly available X-ray projections of industrial products for training and benchmarking DCNNs. A notable exception is GDX-ray \cite{mery2015gdxray} - a dataset containing projections of welds, castings, baggage, and natural objects. Data for a specific defect detection problem have to either be acquired manually or generated computationally. There are two main approaches to data generation: transforming real-world projections with image-to-image methods and simulating X-ray imaging. Image-to-image algorithms can be used to add defects to existing images of objects without defects \cite{van2022inline}. Generative Adversarial Networks (GANs) can also be used to create new data similar to the real data \cite{tempelaere2023synthetic} or to perform style transfer from one imaging modality to another \cite{armanious2020medgan}.

\subsection{Simulation of X-ray imaging and the effect of scattering}

Simulation of X-ray imaging requires knowledge of the experimental setup and the 3D structure of the studied object. There are two categories of simulation algorithms: probabilistic and deterministic. The \textit{probabilistic} approach uses Monte-Carlo (MC) methods to imitate the stochastic nature of real X-ray interactions with matter, and can produce highly accurate results. The particle physics toolkit GEANT4 \cite{agostinelli2003geant4} is used as a gold standard to verify other algorithms. The GEANT4-based software GATE is often used to simulate different modalities of medical imaging (PET, SPECT, CT) and dosimetry \cite{jan2011gate}. The main drawback of these methods is the computational cost. The generated projections contain stochastic noise that can only be reduced by simulating a large number of X-ray photons. The \textit{deterministic} approach uses ray-tracing algorithms to compute projections faster by using a simplified model of X-ray interactions. aRTist \cite{bellon2007artist} (Analytical RT Inspection Simulation Tool) was developed as a fast simulation software to generate realistic X-ray projections based on the mesh model of the object. A similar approach was proposed for baggage inspection \cite{gong2018rapid} using GPU-based ray-tracing. It was later shown \cite{gong2019rapid} that X-ray scattering can also be simulated with additional computational cost. 

Simulation of X-ray photon scattering is missing from many probabilistic algorithms. When it is implemented, it increases the computational cost by orders of magnitude \cite{gong2019rapid}. A faster alternative is to approximate the distribution of scattered photons; in particular using convolutions \cite{sun2010improved, bhatia2016scattering}. Convolution kernels can be extracted from Monte-Carlo simulations or from experimental measurements. However, these methods are limited by the difficulty of finding a small number of kernel parameters that work for various objects. Deep learning scattering estimation \cite{maier2018deep} has been proposed as an alternative to kernel-based methods, but it requires a larger amount of data with known scattering patterns.

If the scattering effect is not reduced experimentally or compensated algorithmically, the projection quality may be compromised \cite{ruhrnschopf2011general}. This problem is thoroughly studied in radiology, where scattering can reduce the contrast in projections, making them unsuitable for diagnostic purposes \cite{barnes1991contrast}. The amount of scattered radiation is usually measured by calculating the scattering-to-primary ratio (SPR). The ratio depends on the field of view, the air gap between the patient and the detector, and X-ray energy \cite{cardoso2009evaluation}. SPR can be reduced by hardware techniques, such as anti-scatter grids. There are several metrics that quantify the change in SPR: Contrast Improvement Factor (CIF), Bucky Factor (increase in absorbed dose), and change in signal difference to noise ratio SdNR \cite{cunha2010evaluation}.

Many important insights can be found in radiology articles on the connection between scattering, image quality, and quality for diagnostic purposes. The effect of scattering on some image properties, such as the point spread function, can be calculated \cite{boone1988analytical}. Using clinical trials, it is possible to infer the effect of image properties such as noise power spectrum, resolution, and point spread function on image quality for diagnostic purposes \cite{martin1999measurement, jessen2004balancing}. To our knowledge, there have been no studies correlating the presence of scattering with a decrease in diagnostic accuracy that would allow a quantitative evaluation of the impact of anti-scatter techniques.

\section{Methods}

We formulate the defect detection task as a segmentation problem. The goal is to compute a labeled image $y \in \mathcal{Y}$, where each pixel has a label according to its classification, from the measured X-ray projection $x \in \mathcal{X} \subset \mathbb{R}^m$. The set of possible labels $\{0,...,k\}$ corresponds to the defect types and depends on the particular detection task. An algorithm performing defect detection is described as

\begin{equation}
\label{alg_def}
    f_\phi: \mathcal{X} \rightarrow \mathcal{Y},
\end{equation}

where $\phi$ are the parameters of the algorithm. Supervised deep learning considers hypothesis functions $f_\phi \in \mathcal{H}$ with a high-dimensional parameter space and determines $\phi$ by solving the optimization problem. 

\begin{equation}
    \phi = \argmin_{\phi} \sum_{i=1}^{N_{train}} l(f_\phi(x_i), y_i), (x_i, y_i) \in S_{train},
\end{equation}

where $S_{train} = \{(x_i, y_i)\}_{i=1}^{N_{train}}$ is a training dataset consisting of projections and correct predictions (ground-truth) and $l$ is a loss function. By including and excluding the scattering signal, we create two projections $x_i^{MC}$ and $x_i^{R}$ that correspond to the same object and have the same ground-truth $y_i$. Consequently, these two types of projections form two versions of the training dataset $S_{train}^{MC}$ and $S_{train}^{R}$. Since these sets are not equal, they lead to the different solutions $\phi^{MC}$ and $\phi^R$ of the optimization problem.

Due to the low interpretability of deep learning methods, the accuracy of $f_\phi$ is usually characterized by applying the function to the test set $S_{test} = \{(x_i, y_i)\}_{i=1}^{N_{test}}$. With an accuracy metric $h: \mathcal{Y} \times \mathcal{Y} \rightarrow \mathbb{R}$, the accuracy of the function with parameters $\phi$ is characterized by the sequence

\begin{equation}
    A^\phi = \{h(f_\phi(x_i), y_i)\}_{i=1}^{N_{test}}
\end{equation}

We propose to analyze $A^\phi$ using POD curves, which can be determined by computing the regression between the accuracy $A_i^\phi$ for each object and the vector of corresponding properties of the object and projection $\textbf{t}_i$ (vector of independent variables). A simple example of the object property correlated with the detection accuracy is the defect size (later referred to as $s_i$), and the corresponding vector $\textbf{t}_i = \left(\begin{smallmatrix}1 & s_i\end{smallmatrix}\right)$
We assume that with a sufficiently good choice of properties, $\textbf{t}_i$ is correlated with the probability of detection - the probability $P$ that the value of $A_i^\phi$ exceeds the threshold $A_{thr}$. The regression is computed using a generalized linear model according to the equation

\begin{equation}
\label{regression_eq}
    g(P(A_i^\phi > A_{thr})) = \textbf{t}_i \pmb{\beta},
\end{equation}

where $g: \mathbb{R} \rightarrow \mathbb{R}$ is a link function (defining the shape of the POD curve) and $\pmb{\beta}$ is a vector of regression coefficients. Then $\pmb{\beta}$ can be used to characterize the performance of the algorithm with parameters $\phi$ in a data-driven way. For a desired accuracy $A_{thr}$ achievable with probability $P_{good}$, a set of constraints for $\textbf{t}$ can be computed with a confidence interval given by the regression uncertainty. By comparing the thresholds of $\textbf{t}$ we determine whether the difference between $\phi^{MC}$ and $\phi^R$ is significant with respect to the object properties. Consequently, this answers the question whether $S_{train}^{MC}$ can be replaced by $S_{train}^{R}$ without losing detection accuracy.

\begin{figure}[ht]
%\begin{figure}[p]
\centering
\includegraphics[width=0.60\linewidth]{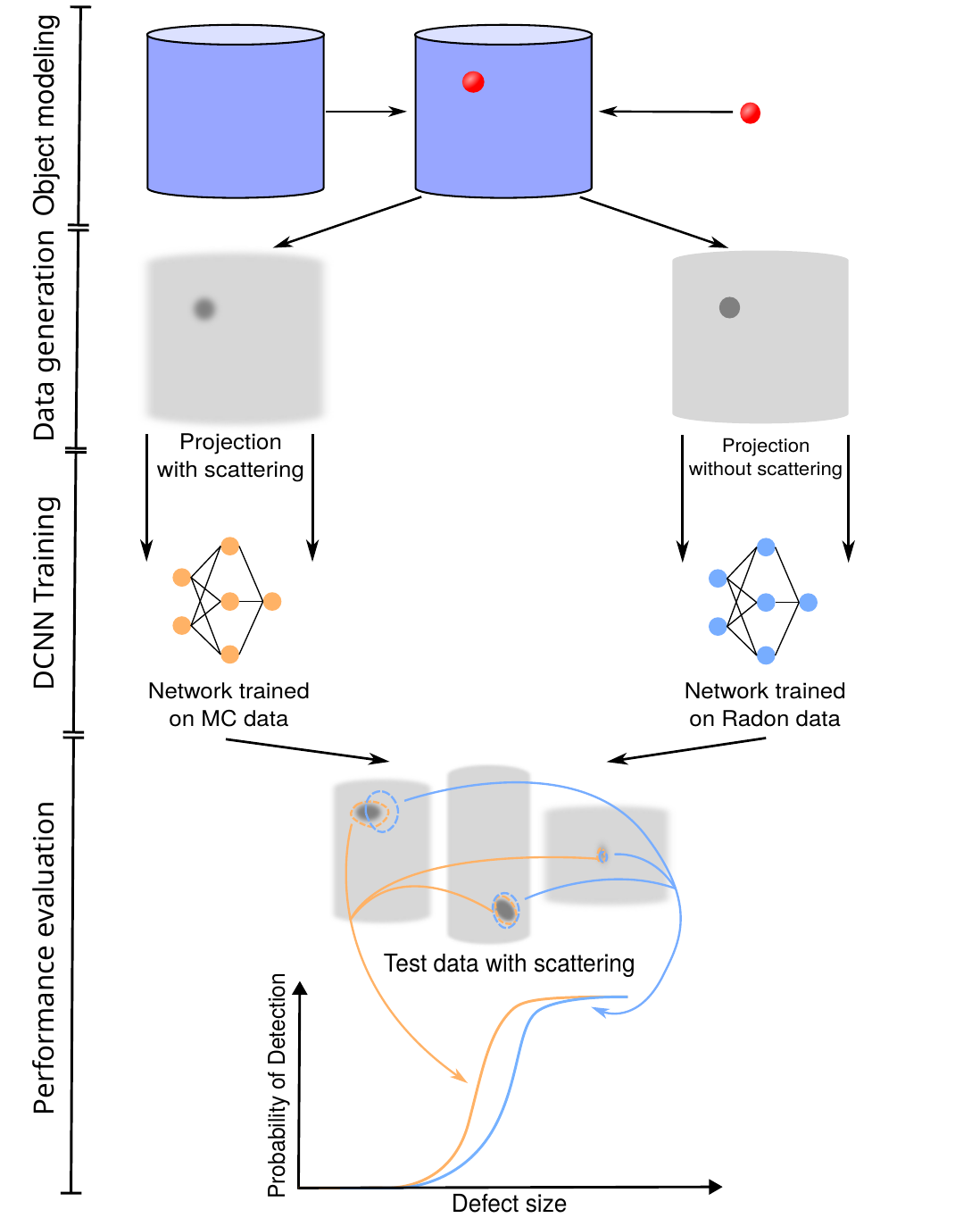}
\caption{Application-driven approach to evaluate the difference between simulation with and without scattering. First, a large number of 3D volumes is generated by combining different variations of object and defect geometry. Two forward projection methods are used to transform the 3D volumes into X-ray projections. Each dataset containing a variety of projections is used to train a DCNN. They are applied to the same collection of test data with scattering. The performance of the DCNNs is evaluated using POD curves.}
\label{methodology}
\end{figure}

We test this methodology on generated data following the approach formulated in our previous paper \cite{andriiashen2023ct} (Fig. \ref{methodology}). The data are generated based on a collection of 3D volumes that define the physical properties of the objects independent of the imaging method. Two algorithms (including and excluding scattering) are used to transform the volumes into projections $x_i^{MC}$ and $x_i^{R}$ by simulating X-ray imaging. These projections are combined into datasets $S_{train}^{MC}$ and $S_{train}^{R}$ to train DCNNs $f_{\phi^{MC}}$ and $f_{\phi^{R}}$. Both networks are tested on the same dataset $S_{test}^{MC}$. The POD analysis is performed to find a correlation between $A^\phi$ and defect size. Thus, we can compare the smallest defect size detectable with $\phi^{MC}$ and $\phi^{R}$ and conclude if the difference is statistically significant.

\subsection{Object modeling}

Studying the influence of scattering imposes several constraints on the data generation approach. In our methodology, we want to avoid cases where a DCNN fails to detect a defect because it was not exposed to a certain morphology during training. This problem is well-known in deep learning as Out-of-Distribution performance \cite{hendrycks2017a, hendrycks2021many}. For example, if a network was trained to detect defects in large objects, it might fail on a test projection with a small object, regardless of scattering. The objects used for the training and test sets should have the same distribution over $\mathcal{X}$, and as many objects as possible should be in both sets. In real-world applications, test objects are measured experimentally, while training objects are created digitally. Thus, it is difficult to guarantee that the distributions of training and test objects match.

We address these limitations by choosing to work with a model problem. Each 3D volume is created by combining a small number of objects according to a parameterizable algorithm. Training and test sets of volumes are created with different random sequences of parameters, each parameter having the same distribution for both sets. We assume that a large enough number of volumes is generated, so that the training and test datasets have similar distributions over $\mathcal{X}$. We check this assumption qualitatively by visual inspection of the detection performance, but a detailed evaluation is beyond the scope of this paper. In addition to the geometry, the volume is characterized by the material properties. To consider a variety of scattering patterns, we repeat the experiments with the same geometry and different material compositions.

The model problem we use in this study is the detection of cavities in cylinders. The object is a homogeneous cylinder parameterized by radius and height. The defect is an ellipsoidal void inside the cylinder. It is defined by ellipsoidal axes, height, and distance from the rotational axis of the cylinder. Materials of the cylinder are chosen from the common materials of industrial products: PMMA (type of plastic, synthetic polymer $(C_5 O_2 H_8)_n$), aluminum, and iron.

\subsection{Data generation}

To ensure that the scattering distribution in the generated data is accurate and similar to real measurements, we use Monte-Carlo simulation as implemented in the GATE framework \cite{jan2011gate}. An accurate Monte-Carlo simulation requires a detailed knowledge of the experimental setup: X-ray source, studied object, and detector. However, not all of these are crucial for an accurate scattering simulation. The most important properties are the energy spectrum of the emitted X-ray photons, the attenuation curves of the materials present in the object, and an accurate description of the object geometry (defined as a voxelized volume, a mesh, or a collection of simple shapes). GATE simulates each emitted photon individually and records its coordinates if it is registered by the detector.

If the simulation considers only X-ray absorption, the number of photons detected in a pixel $p$ is given by
\begin{equation}
\label{abs_photons}
    I_{p}^{abs}(E) = I_0(E) \exp{\left( -\int_{l_p} \mu(E, x) dx \right)},
\end{equation}
where $\mu(E, x)$ is a 3D distribution of the object's attenuation coefficient for the X-ray energy $E$, $l_p$ is the trajectory from the X-ray source to the pixel $p$.  The acquired X-ray projection $x^R$ is computed by summing the photons of all energies and correcting for the flatfield (the projection acquired without the object)

\begin{equation}
\label{ff_cor}
    x_p^R = -\log \frac{\int_E  I_p^{abs}(E) dE}{\int_E I_0(E)dE}.
\end{equation}

This formulation does not include the effect of the detector gain \cite{whiting2002signal}. When X-ray scattering is simulated, the number of photons detected $I_p$ is $I_p^{abs} + I_p^{scat}$. The number of scattered photons $I_p^{scat}$ is not defined by the trajectory $l_p$. Instead, each voxel of the object $v$ contributes to the amount of scattering according to the equation
\begin{equation}
\label{scat_photons}
    dI_{pv}^{scat}(E') = \frac{1}{\sigma_v(E)} \frac{d\sigma(\theta_{pv}, E)}{d\Omega} I_v(E) d\Omega_{pv},
\end{equation}
where $\frac{d\sigma}{d\Omega}$ is the differential scattering cross-section, $\sigma_v(E)$ is the total cross-section for all scattering from the voxel $v$, $\theta_{pv}$ is the scattering angle, $I_v(E)$ is the energy spectrum of the photons in the voxel, and $d\Omega_{pv}$ is the solid angle corresponding to $p$ seen from $v$ \cite{freud2005hybrid}.
Differential cross-sections of Rayleigh and Compton scattering as a function of scattering angle and energy are known. Simulation of scattering makes the signal in a pixel dependent not only on a single trajectory but on the entire object. Furthermore, multiple scattering is possible, and $I_v(E)$ includes photons scattered from other parts of the object. With a sufficiently large number of simulated photons, Monte-Carlo methods consider many possible scattering trajectories and compute the total number of scattered photons $I^{scat}$. Similar to Eq. \ref{ff_cor}, raw measurements should be flatfield corrected to compute the projection with scattering

\begin{equation}
\label{ff_cor_mc}
    x_p^{MC} = -\log \frac{\int_E \left[ I_p^{abs}(E) + I_p^{scat}(E) \right] dE}{\int_E I_0(E)dE}.
\end{equation}

The model problem simplifies the definition of object geometry and material composition. The object geometry is defined as a mixture of a cylinder and an ellipsoid which are the basic object shapes for GATE. While any object can be simulated based on its polygonal mesh, the computational cost of photon tracking increases with object complexity. Limiting object geometry to basic shapes accelerates the simulation. The attenuation properties are automatically calculated for a given chemical formula. We compute the X-ray source spectrum with an empirical model of an X-ray tube (implemented in xpecgen \cite{hernandez2016xpecgen}), which provides an energy spectrum depending on the voltage. Several voltage values are chosen for each material, since the scattering distribution depends strongly on both properties. 

An image produced by a Monte-Carlo simulation is inherently noisy due to the limited number of generated photons. It can be shown that the number of registered photons follows Poisson distribution \cite{whiting2002signal} (Compound Poisson for energy-integrating detectors). Noise in real X-ray projections is additionally influenced by other factors, such as noise in the detector electronics \cite{whiting2006properties}, focal spot size, and scintillator blur \cite{smalley2020image}.
We choose to perform a simulation with a perfect detector that registers every photon reaching it (technically implemented as a sufficiently thick layer of a heavy material). Simulation of a realistic detector is possible with Monte-Carlo methods but is beyond the scope of this study. Such modeling would require detailed knowledge of a real detector, which is often lacking in industrial tasks.

\begin{figure}[ht]
%\begin{figure}[p]
\centering
\includegraphics[width=0.9\linewidth]{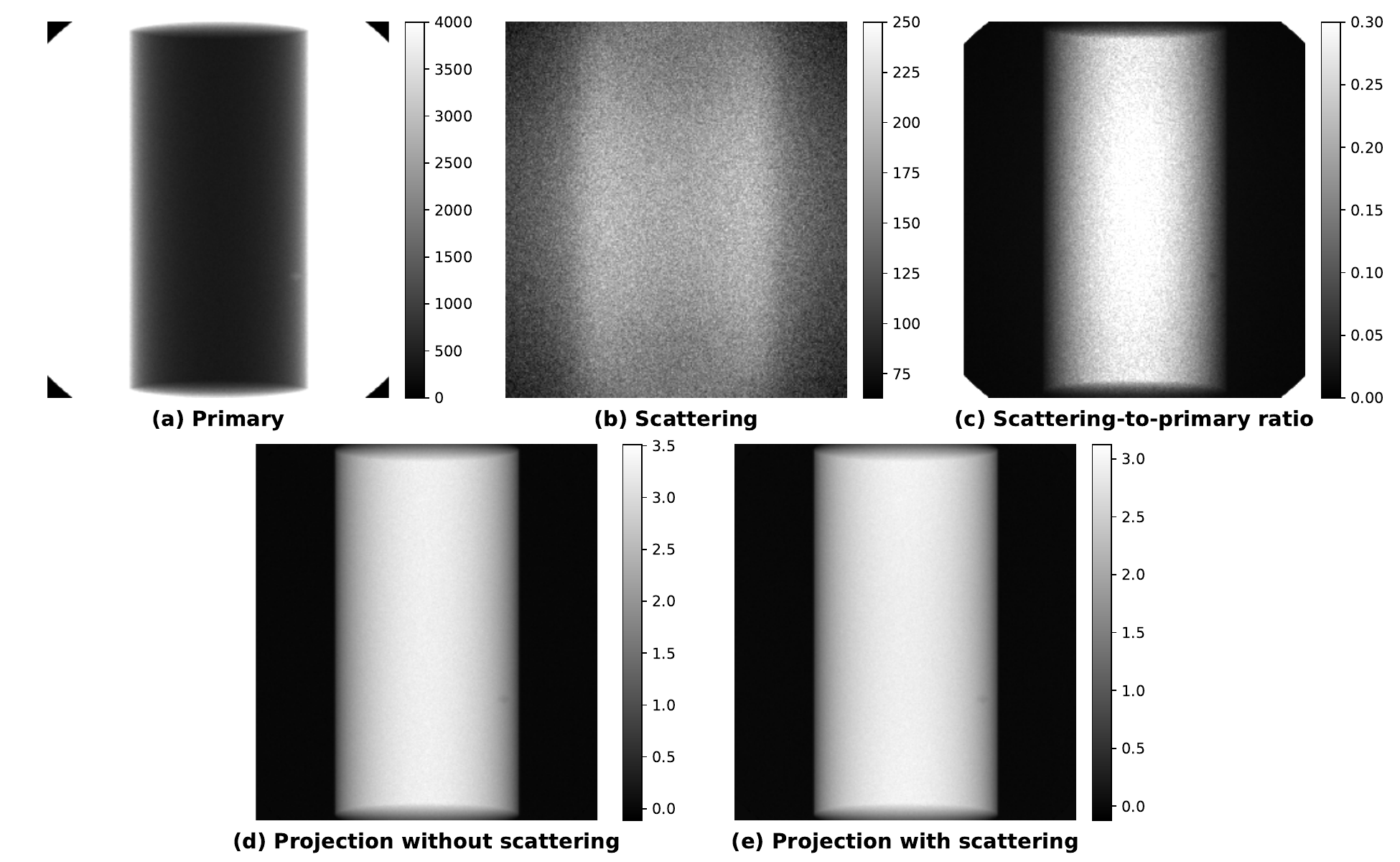}
\caption{Different distributions of the X-ray signal that are computed with a Monte-Carlo algorithm: distribution of primary photons registered by the detector (a), distribution of scattered photons (b), scattering-to-primary ratio (c), projection image without scattering after pre-processing (d), projection image with scattering (e).}
\label{singleim_mc}
\end{figure}

The output of GATE includes the coordinates where each X-ray photon was detected and how many times it scattered before detection. After matching the coordinates with different pixels of the detector, a total distribution of primary photons of all energies $I^{abs}$ (Fig. \ref{singleim_mc}a) and a distribution of scattered photons $I^{scat}$ (Fig. \ref{singleim_mc}b) integrated over energy and all possible scattering centers can be calculated. The distribution of SPR is defined as $\frac{I^{scat}}{I^{abs}}$ (Fig. \ref{singleim_mc}c). Following Eq. \ref{ff_cor} and \ref{ff_cor_mc}, distributions of detected photons are converted into pre-processed projections with and without the scattering signal (Figs. \ref{singleim_mc}e and \ref{singleim_mc}d).

\subsection{DCNN Training}

We train segmentation DCNNs to perform defect detection. This approach not only detects the presence of the defect but also provides its location on the projection. The training process requires input data (previously described $x^R$ and $x^{MC}$) and the ground-truth with defect locations. To construct the ground-truth, the Radon transform is applied to 3D volumes. The result of this Radon transform is a 2D distribution of material length along the ray from the source to the detector. This image is used to outline the location of the defect and compute its geometric properties. The DCNN trained on projections that include scattered photons is referred to as the network trained on MC data. The one trained on data without the scattering signal is referred to as the network trained on Radon data.

DCNN training as an optimization problem for network weights $\phi$ is affected by random factors (initialization, GPU computation). The performance of a single network is not sufficient to draw conclusions about the influence of training data. Therefore, each training is repeated multiple times to have multiple instances $\phi$ of the same DCNN. The performance of the instances is then averaged to determine the performance of the DCNN.

The DCNN architecture determines the structure of the weights $\phi$ and has a significant effect on the performance. Before selecting an architecture for a majority of the experiments, we performed one analysis with multiple state-of-the-art segmentation architectures (the results are given in the Appendix). It was observed that different architectures with commonly used metaparameters converge to a similar level of accuracy. Thus, the influence of scattering could be studied for one architecture, and similar results are expected for others.

We choose the Mixed-Scale Dense Convolutional Neural Network (MSD) \cite{pelt2018mixed} for a majority of experiments. This architecture is an alternative to classical convolutional neural networks with scaling, and is aimed towards easy training and reusability. The MSD network consists of multiple layers with the same set of operations (convolution, summation and activation). All layers produce feature maps of the same dimensions and are connected to each other (dense connections). Dilated convolutions are used to capture image features at different scales (mixed scale). Due to dense connections and a small number of parameters (compared to other popular architectures such as UNet), an MSD network can be trained quickly and provide accurate segmentations. All networks share the same architecture parameters: a depth of 1, a width of 30, dilations in the range from 1 to 10, 1 input channel, and 2 output channels. The total number of weights is close to 5000. We have tested the performance with different values of width and dilations and have not observed a noticeable improvement compared to the standard parameters. The optimization of weights is done by ADAM optimizer with a learning rate of $10^{-3}$. Network training and testing is done in PyTorch framework with MSD implementation from \cite{hendriksen-2019-msd-pytor}.

\subsection{Performance evaluation}

Many metrics $h$ can be applied to evaluate the similarity between the output of the segmentation DCNN and the ground-truth. We use the $F_1$ score, which can be calculated according to the equation
\begin{equation}
    F_1 = \frac{2TP}{2TP + FP + FN}
\end{equation}
where TP (True Positive) is the number of defect pixels that were correctly segmented, FP (False Positive) is the number of pixels that were incorrectly marked as corresponding to the defect, and FN (False Negative) is the number of defect pixels that were missed by a segmentation algorithm.

\begin{figure}[ht]
%\begin{figure}[p]
\centering
\includegraphics[width=0.99\linewidth]{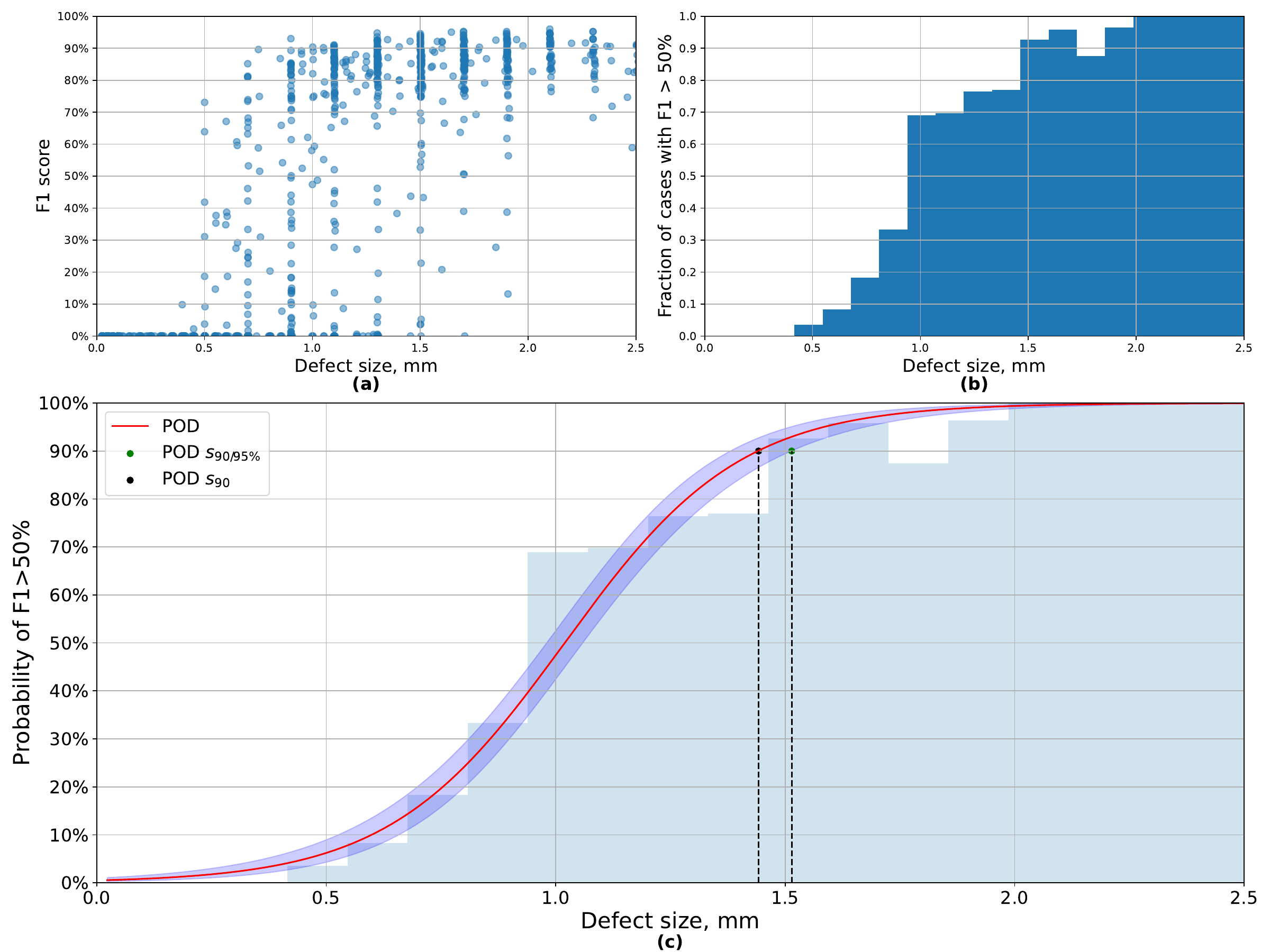}
\caption{Example of a POD curve and its relation to accuracy for a single projection: (a) the correlation between $F_1$ score and defect size for all projections in the test set, (b) a histogram computed after binning the defect size indicating that the fraction of projections with $F_1>50\%$ increases with defect size, (c) the POD curve representing a probability of $F_1>50\%$ with the smallest detectable defect $s_{90}$ and its higher bound highlighted.}
\label{pod_pipeline}
\end{figure}

Fig. \ref{pod_pipeline}a illustrates that $F_1$ varies significantly between different test cases with different defect sizes. By size, we refer to the largest intersection between the defect and the primary X-ray trajectories (the largest thickness). This is not the only way to characterize the defect. For example, its area or perimeter can also be used to represent size. These properties are often correlated because expected defects have similar shapes. Following the POD methodology, we convert the segmentation accuracy into a binary variable by the defining segmentation as successful if $F_1 > 50\%$. In Fig. \ref{pod_pipeline}b, the projections are divided into groups by defect size binning and a fraction of successful segmentations is computed for each bin. Segmentation accuracy improves as the defect size increases.

The properties of the POD curve are determined by computing the regression according to Eq.\ref{regression_eq}, where \textbf{t} is a constant and the defect size. The link function $g$ defines the shape of the POD curve and reflects the dependence between the probability of successful segmentation and the defect size. The probability is close to zero for small defects, one for large defects, and there is a transition in between. The generalized linear model in this case is represented by the equation

\begin{equation}
    g(P(F_1 > 50\%)) = \ln \frac{P(F_1 > 50\%)}{1-P(F_1 > 50\%)} = a + b s,
    \label{glm_eq}
\end{equation}

where $a$ and $b$ are the fit parameters. Their values are computed by solving an optimization problem of maximizing likelihood of getting the observed detectability outcomes for the test set with known values of defect size. The logit function $g(x) = \ln \frac{x}{1-x}$ can be replaced by other functions such as log-log, complementary log-log, probit, etc. An example of a POD curve is shown in Fig. \ref{pod_pipeline}c (computed using the statsmodels package \cite{seabold2010statsmodels}). 

The parameters $a$ and $b$ describe the performance of the algorithm $f_\phi$. To conclude whether two algorithms have the same performance in this context, the values of $a$ and $b$ should be compared, taking into account the uncertainty interval. In POD analysis, it is common to approach this problem in a task-driven way and compute the value of $s$ leading to a fixed level of good detection probability (usually 90\%) according to the equation

\begin{equation}
    s_{90} = \frac{g(90\%) - a}{b}
\end{equation}

This value $s_{90}$ describes the size of the smallest detectable defect. Due to regression uncertainty, the true value of $s_{90}$ lies in a region around the computed value. Thus, if two sets of parameters $\phi^R$ and $\phi^{MC}$ have different values of $s_{90}$, there is a probability that the true value is the same in both cases, and the difference is only caused by the fit uncertainty. We address this problem by computing an upper bound for $s_{90}$ which is referred to as $s_{90/95}$. The true value of $s_{90}$ is less than $s_{90/95}$ with 95\% probability. Therefore, if the computed value of $s_{90}$ for $\phi^R$ is less than $s_{90/95}$ for $\phi^{MC}$, the difference in performance is statistically significant. Otherwise, it can be explained by the fit uncertainty with less than 95\% probability.

The accuracy of segmentation is expected to depend on many projection properties. In particular, the defect location and surrounding object features are also used in POD analysis \cite{chen2021multivariate}. When the POD curve is computed, the performance for a particular value of defect size is averaged over different objects, different defect locations and their shapes. More projection properties could be included in a multivariate POD curve to produce a more detailed description of detection performance. We use a multivariate fit to study the influence of SPR on detection. The corresponding generalized linear model is given by the equation

\begin{equation}
    g(P(F_1 > 50\%)) = a + b s + c SPR.
    \label{multivariate_fit}
\end{equation}

\subsection{Implementation details}

\begin{figure}[ht]
%\begin{figure}[p]
\centering
\includegraphics[width=0.9\linewidth]{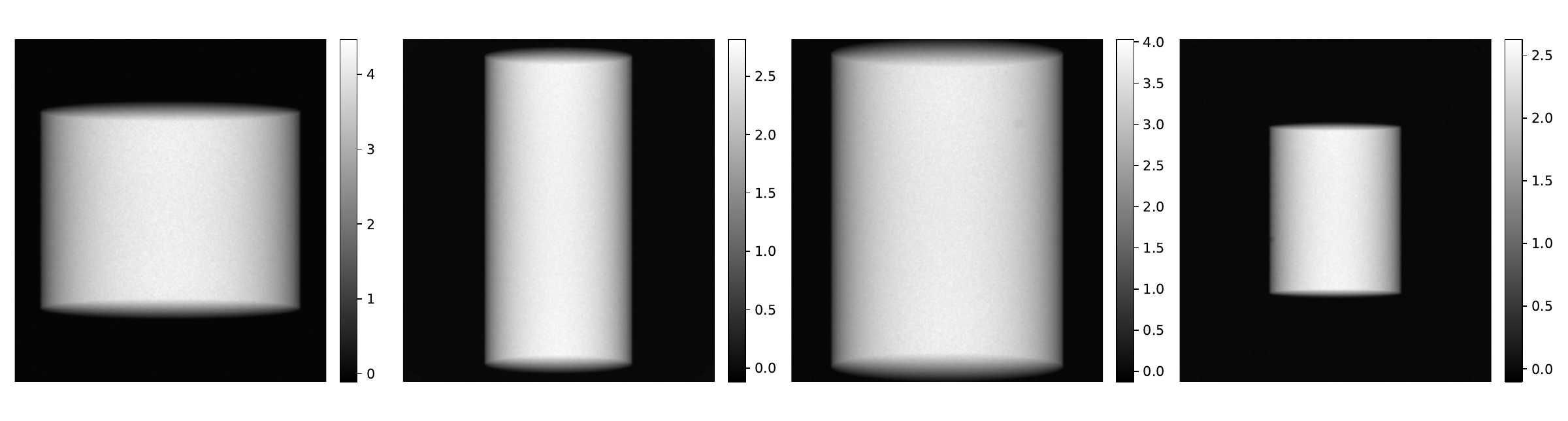}
\caption{Examples of projections corresponding to different generated volumes.}
\label{projection_ex}
\end{figure}

The geometric properties of the objects for the training and test sets are generated in advance and kept the same for all materials and voltages. We generate 1250 objects for the training set (1000 for training, 250 for validation) and 1000 objects for the test set. The radius of the cylinder ranges from 1 to 25~mm and the height varies from 20 to 55~mm. Ellipsoidal cavities are generated as deformed spheres: first, a radius is chosen to be in the range from 0.1~mm to 1~mm, and then the ratio between each axis and this radius is in the range from 0.7 to 1.3. A cavity can be placed at any distance from the axis of rotation. The acquisition geometry remains the same for all datasets. The source-object distance (SOD) is 200 mm and the source-detector distance (SDD) is 300 mm, resulting in a magnification factor of 1.5. The detector plane is a 75$\times$ 82.5 $\textrm{mm}^2$ rectangle with a pixel size of 0.3~mm. Thus, the generated projections have a size of 250 $\times$ 275 $\textrm{px}^2$. One projection contains MC simulations of $10^9$ emitted photons. Examples of these projections are shown in Fig. \ref{projection_ex}.

\begin{figure}[ht]
%\begin{figure}[p]
\centering
\includegraphics[width=0.99\linewidth]{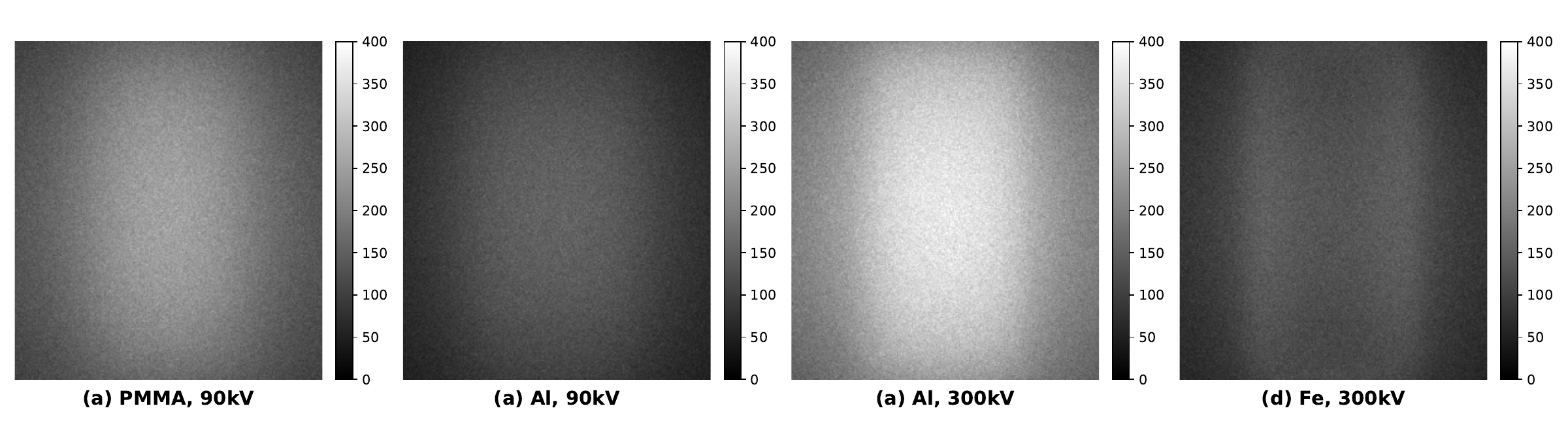}
\caption{Comparison of scattering distributions for the same object made of different materials and inspected with different tube voltages: (a) PMMA at 90~kV, (b) aluminum at 90~kV, (c) aluminum at 300~kV, (d) iron at 300~kV. For the same voltage, the number of scattered photons is higher for materials with lower atomic numbers. For the same material, the amount of scattering is higher for higher voltage. Iron objects have a different scattering pattern with more photons near the edges.}
\label{mat_scat_comp}
\end{figure}

Different material compositions and voltages are used to explore a variety of scattering distributions. In practice, heavier materials are inspected at higher voltages. Otherwise, only a small fraction of the X-ray radiation penetrates the object, resulting in a high noise level. The combination of material and voltage can be characterized by the value of HVL (Half-Value Layer) - the thickness of the object at which the intensity of X-ray entering it is reduced by half. PMMA is simulated at 90~kV (HVL = 27.8~mm) and 150~kV (HVL = 31.7~mm). Aluminum is simulated at 90~kV (HVL = 4.28~mm), 150~kV (HVL = 6.25~mm) and 300~kV (HVL = 17.9~mm). Iron is simulated at 300~kV (HVL = 3.85~mm) and 450~kV (HVL = 4.6~mm). Fig. \ref{mat_scat_comp} shows different scattering distributions for the same object depending on material and voltage. The number of scattered photons (not necessarily SPR) increases with higher voltage and lower atomic number. Furthermore, there is a qualitative change in the spatial properties: for materials with low atomic number scattering is more uniform, and for heavier materials it follows the shape of the object. This can be explained by different ratios of Rayleigh and Compton scattering and probabilities of further absorption in the object for scattered photons.

\section{Results}

\subsection{Comparison of DCNN performance}

For each combination of material and voltage, two DCNNs with the same architecture are trained on data without scattering (Radon data) and with scattering (MC data), 10 instances each. Averaging the segmentation accuracy of the DCNN on the test dataset does not lead to a good performance estimate due to the high variance. For example, in the case of iron at 450~kV, the network trained on Radon data has a segmentation accuracy of 43\% $\pm$ 40\% on the test dataset, while the network trained on MC data achieves 44\% $\pm$ 40\%. A high variance is observed for each dataset, making a direct comparison of the DCNNs impossible.

\begin{figure}[ht]
%\begin{figure}[p]
\centering
\includegraphics[width=0.99\linewidth]{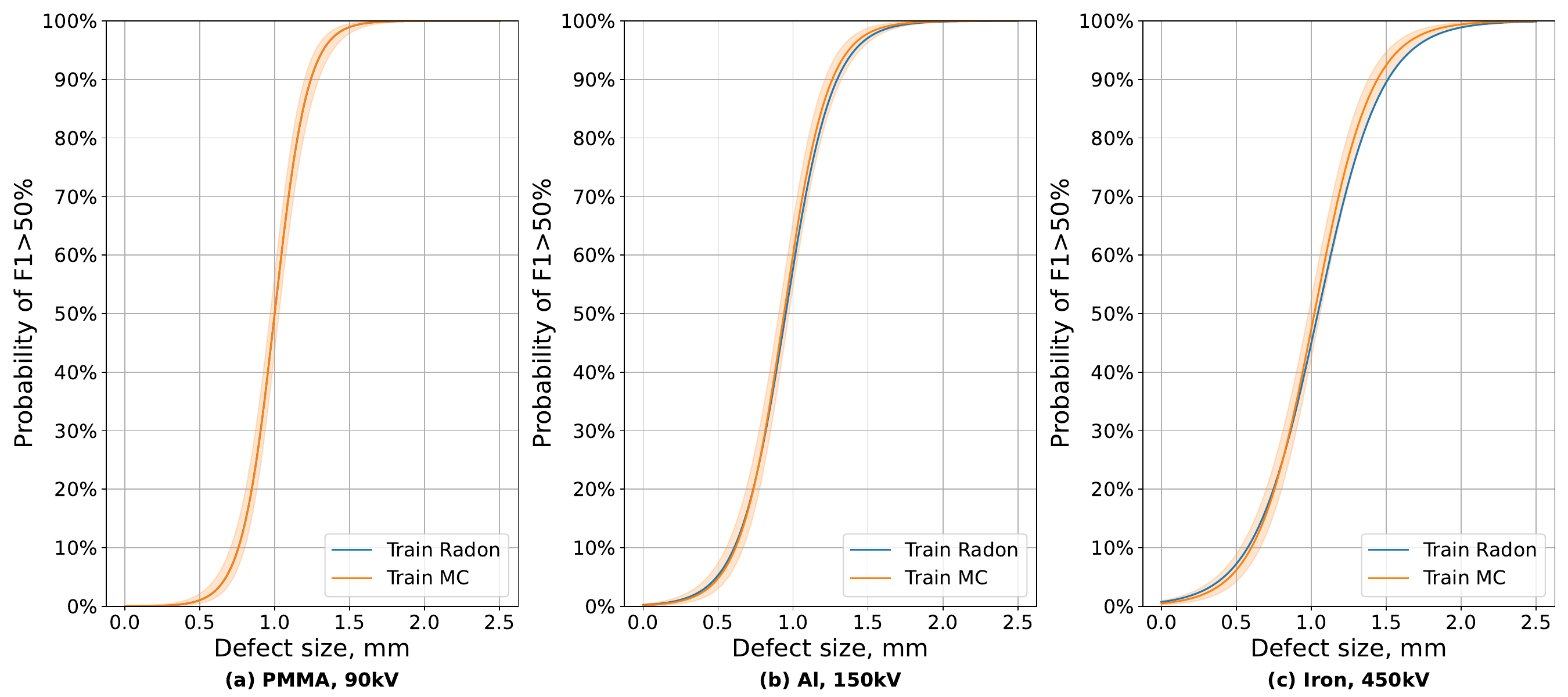}
\caption{Comparison of POD curves for different datasets: (a) PMMA at 90~kV, (b) aluminum at 150~kV, (c) iron at 450~kV. The difference between training on projections with and without scattering is negligible for PMMA. It becomes larger for aluminum but within a 95\% confidence interval. For iron at 450~kV, the network trained on MC data performs better than the network trained on Radon data considering the confidence interval.}
\label{comp_mat}
\end{figure}

For a detailed analysis of the performance, POD curves are computed for both DCNNs for each voltage and material combination. Fig. \ref{comp_mat}a shows a pair of POD curves for PMMA at 90~kV with no difference in performance between the networks trained on Radon and MC data. The POD curve for the Radon data network lies within the 95\% confidence interval of the MC POD curve, making the difference between them statistically insignificant. The opposite case is shown for iron at 450~kV (Fig. \ref{comp_mat}c) where higher POD values for the network trained on Radon data lie below the confidence interval for the network trained on MC data (in particular, $s_{90}$). Hence, there is a statistically significant difference between the training sets, and MC data lead to better performance. Nevertheless, the difference between $s_{90}$ is only 5\% for iron at 450~kV and smaller for other datasets (e.g. for aluminum at 150~kV as shown in Fig. \ref{comp_mat}b). Furthermore, for defects larger than $s_{90}$, the average segmentation accuracy is 82\%$\pm$16\% for the network trained on MC data and 82\%$\pm$18\% for the network trained on Radon data. Although the variance is smaller than for the whole dataset, the performance of the two networks is still effectively the same.

The values of $s_{90}$ for both networks in all cases are shown in Table \ref{s90_comp}. In most cases (except for iron at 450~kV) the difference in the smallest segmentable defect between the networks is less than the confidence interval. This serves as an upper estimate for the possible performance difference. Depending on the material and voltage, the relative difference in $s_{90}$ is between 0\% and 3\% and is unlikely to exceed 5\% even if the fit coefficients are not precisely determined. 

\begin{table}[ht]
%\begin{table}[p]
\centering
\begin{tabular}{|l|l|l|l|l|}
\hline
Material, kV      & MC data net, s90 {[}mm{]} & MC data net, s90/95 {[}mm{]} & Radon data net, s90 {[}mm{]} & s90 difference \\ \hline
PMMA, 90~kV       & 1.24                   & 1.29                      & 1.24                      & 0\%            \\ \hline
PMMA, 150~kV      & 1.31                   & 1.36                      & 1.31                      & 0.2\%          \\ \hline
Al, 90~kV         & 1.39                   & 1.46                      & 1.42                      & 1.9\%            \\ \hline
Al, 150~kV        & 1.26                   & 1.32                      & 1.29                      & 2.4\%            \\ \hline
Al, 300~kV        & 1.17                   & 1.22                      & 1.18                      & 0.8\%            \\ \hline
Iron, 300~kV      & 1.59                   & 1.68                      & 1.62                      & 1.7\%            \\ \hline
Iron, 450~kV      & 1.44                   & 1.51                      & 1.51                      & 5.0\%            \\ \hline
\end{tabular}
\caption{Smallest detectable defect size for different materials and training datasets.}
\label{s90_comp}
\end{table}

\subsection{Influence of scattering-to-primary ratio}

\begin{figure}[ht]
%\begin{figure}[p]
\centering
\includegraphics[width=0.99\linewidth]{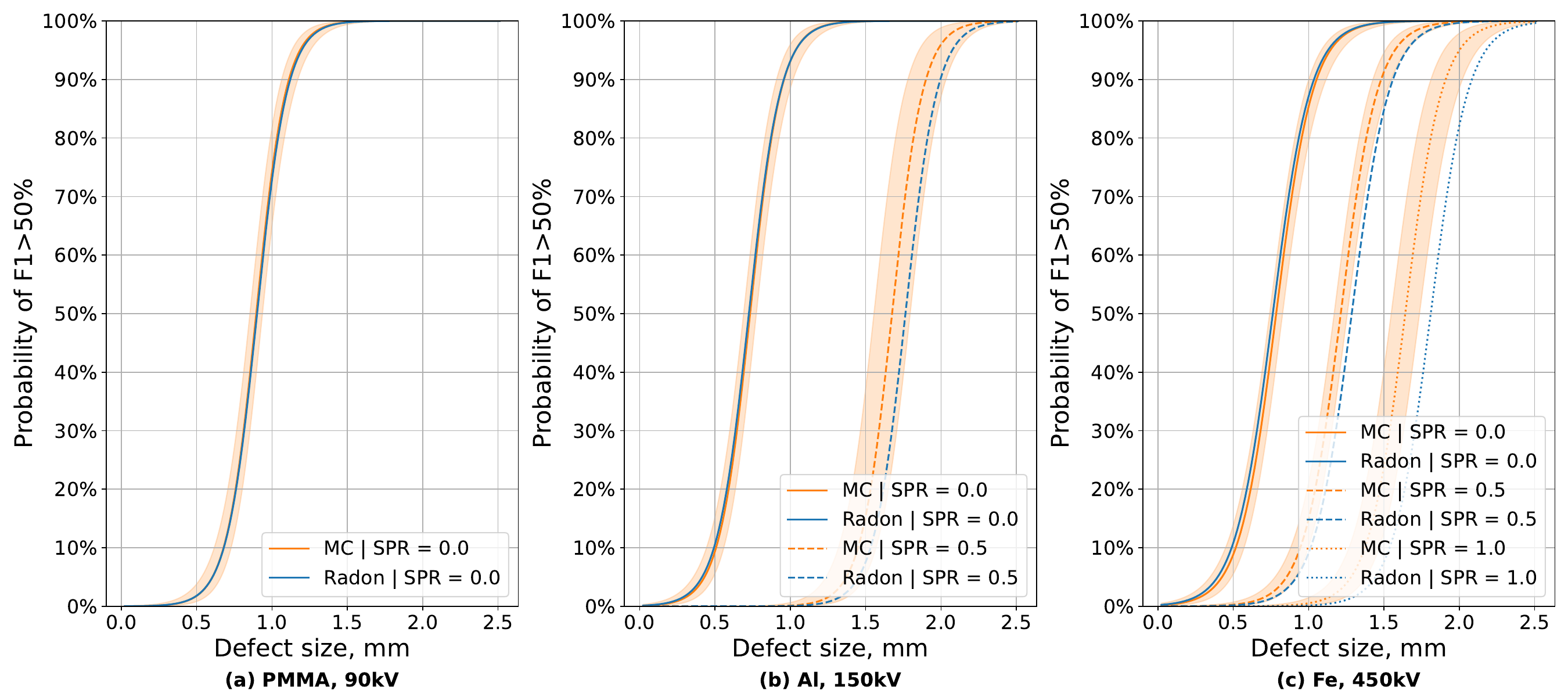}
\caption{Comparison of multivariate POD curves computed at different values of SPR for different datasets: PMMA at 90~kV~(a), aluminum at 150~kV~(b), and iron at 450~kV~(c). Increasing the SPR results in a larger difference between the POD curves of the networks trained on Radon and MC data. Pairs of POD curves are plotted for SPR values of 0., 0.5, and 1.0 if such projections are present in the dataset (the largest SPR for PMMA is 0.16, for aluminum it is 0.55).}
\label{SPR_pod}
\end{figure}

The negligible difference in the performance of the networks trained on Radon and MC data raises the question of biases in the test data and projection properties that influence segmentation. For each dataset, we perform a multivariate POD fit according to Eq. \ref{multivariate_fit} to evaluate the influence of SPR at the defect location (averaged over a region containing the defect). After determining the POD coefficients, the value of SPR can be fixed at a certain level to obtain POD for that amount of scattering signal (Fig. \ref{SPR_pod} for PMMA, aluminum, and iron).

Several observations can be made after splitting the dataset based on the SPR value. First, as the SPR increases, the difference between the networks trained on Radon and MC data also increases. This can be determined either by performing a multivariate fit or by making a univariate fit for a subset of the test data (with a sufficiently high SPR). Second, the values of $s_{90}$ increase as the SPR rises. Third, different datasets have different distributions of SPR. For example, aluminum at 150~kV has 500 objects with SPR $< 0.1$ while iron at 450~kV has 400 such objects. Hence, even if there is no difference between the networks trained on Radon and MC data for low SPR in both cases, the overall POD curve is more affected for aluminum because such cases are more common.

\begin{figure}[ht]
%\begin{figure}[p]
\centering
\includegraphics[width=0.9\linewidth]{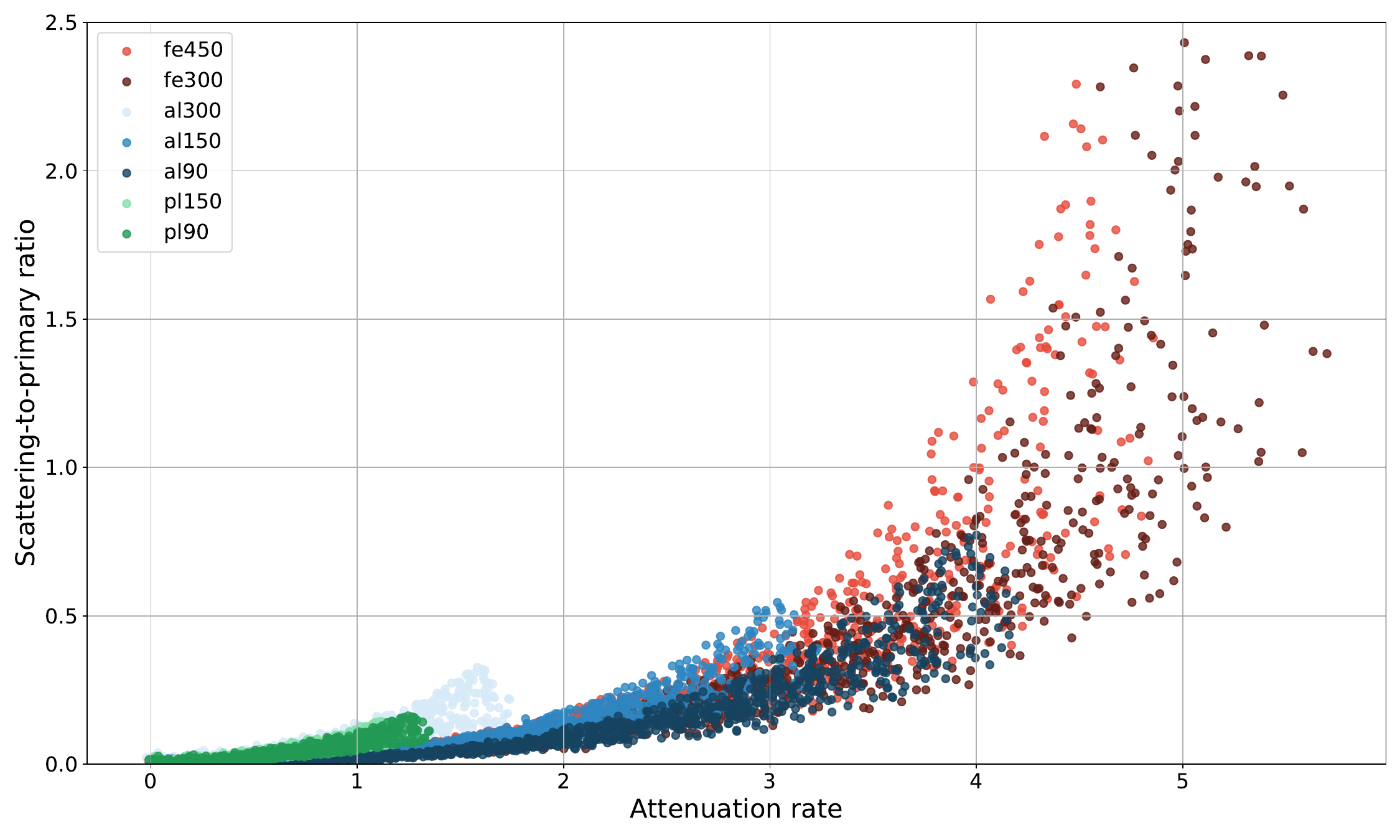}
\caption{Correlation between projection properties at the defect location (SPR and attenuation rate) for different datasets (markers are colored with red for iron, blue for aluminum, and green for PMMA). SPR increases with attenuation rate, and the dependency may be similar for datasets with different materials and voltages.}
\label{scat_att}
\end{figure}

It is important to note that the POD curves only show a correlation between the DCNN accuracy and the different properties of the projection. Furthermore, the value of SPR at the defect location is correlated with other properties such as attenuation rate as shown in Fig. \ref{scat_att}. The attenuation rate $\mu$ in a pixel is calculated as $-\log \frac{I}{I_0}$ where $I_0$ is the total number of photons emitted in the direction of the pixel and $I$ is the number of photons registered (including scattered radiation). The noise in the MC simulation follows a Poisson distribution and depends on $I$. Regions with high SPR often have high $\mu$, leading to high relative noise levels. This correlation may explain the increase in $s_{90}$ for a higher SPR.

In Table \ref{SPR_s90} we compare the $s_{90}$ values in each dataset at the minimum and maximum SPR levels (the minimum is close to 0 in each case). With the sole exception of PMMA at 90~kV (where it can be explained by fit uncertainty), higher SPR leads to a greater difference in performance between the networks trained on Radon and MC data. The geometric structure of the test dataset leads to a large number (approximately 40-50\%) of cases with low SPR, mostly due to the small size of the object. In these cases, there is little difference between training on data with and without scattering. Consequently, the performance on the whole test dataset becomes more similar, even if in other cases the MC network has a smaller $s_{90}$.

\begin{table}[ht]
%\begin{table}[p]
\begin{tabular}{|l|l|ll|ll|}
\hline
\multirow{2}{*}{Material, kV} & \multirow{2}{*}{Max SPR} & \multicolumn{2}{l|}{At SPR = 0}           & \multicolumn{2}{l|}{At Max SPR}           \\ \cline{3-6} 
                              &                          & \multicolumn{1}{l|}{MC, $s_{90}$} & Radon, $s_{90}$ & \multicolumn{1}{l|}{MC, $s_{90}$} & Radon, $s_{90}$ \\ \hline
PMMA, 90~kV              & 0.16                     & \multicolumn{1}{l|}{1.11}    & 1.12       & \multicolumn{1}{l|}{1.59}    & 1.57       \\ \hline
PMMA, 150~kV             & 0.14                     & \multicolumn{1}{l|}{1.22}    & 1.19       & \multicolumn{1}{l|}{1.57}    & 1.63       \\ \hline
Al, 90~kV                & 0.77                     & \multicolumn{1}{l|}{1.00}    & 0.97       & \multicolumn{1}{l|}{2.38}    & 2.51       \\ \hline
Al, 150~kV               & 0.55                     & \multicolumn{1}{l|}{0.96}    & 0.96       & \multicolumn{1}{l|}{1.99}    & 2.09       \\ \hline
Al, 300~kV               & 0.33                     & \multicolumn{1}{l|}{1.07}    & 1.07       & \multicolumn{1}{l|}{1.47}    & 1.50       \\ \hline
Fe, 300~kV               & 5.26                     & \multicolumn{1}{l|}{1.02}    & 1.03       & \multicolumn{1}{l|}{5.84}    & 6.06       \\ \hline
Fe, 450~kV               & 2.29                     & \multicolumn{1}{l|}{1.05}    & 1.04       & \multicolumn{1}{l|}{3.02}    & 3.44       \\ \hline
\end{tabular}
\caption{Difference in $s_{90}$ for minimal and maximal SPR values}
\label{SPR_s90}
\end{table}

\section{Discussion}

We presented a methodology that evaluates the difference between training data for DCNNs using a model problem as an example. This methodology can be applied to various defect detection tasks, as individual parts of the methodology can be replaced to better fit a particular problem. 
The main requirement is the ability to completely remove the X-ray scattering signal for a set of data. It is then possible to compare the dataset containing the scattering signal with the dataset without scattering. For a particular detection task, one should specify an algorithm to perform the task and assemble the test dataset for performance evaluation.
POD analysis is one of many possible ways to evaluate the accuracy of the algorithm. As a more general form of linear regression, POO curves provide a robust technique that can be applied to a wide range of problems without extensive expert knowledge. Depending on the task, a more accurate model relating the DCNN performance to object properties may be suggested.

In the presented model problem, both training and test data are generated using a Monte-Carlo algorithm. This approach provides an easy way to obtain two versions of the same data that differ only in the presence of scattering. Alternatively, the same result can be achieved using the real data and either software or hardware scattering reduction techniques to obtain projections without the scattering signal. The main disadvantage of using experimental data is that the scattering would only be approximately removed, whereas Monte-Carlo simulation can ensure that no scattering is present in the data. The analysis made on simulated data is not necessarily applicable to real measurements, but Monte-Carlo algorithms have been validated to be sufficiently accurate for medical purposes \cite{sarno2018normalized, di2020geant4}. While MC methods can be applied to any problem, objects with complicated morphology and high variance present a computational challenge to the proposed methodology. The computational cost of a single photon simulation depends on the geometric representation and might increase by orders of magnitude for a detailed polygonal mesh. Thus, optimizing the level of detail in a model is crucial for real-world applications. Furthermore, high variability leads to a large number of objects that have to be simulated to produce a representative training set.

Accurate MC simulation requires extensive knowledge of the modeled X-ray system to account for all possible experimental effects. Our model problem does not represent a specific setup, and a number of simplifications has been made. We have not included detector noise in the MC data, although it could be implemented as an image post-processing. According to the calibration of our X-ray system at 90~kV \cite{coban2020explorative}, $10^9$ simulated photons correspond approximately to a high exposure of 1~s at 45~W of tube power. In this case, the Poisson noise component simulated by MC methods is the main source of noise, and the electronic noise of our detector could be ignored. Furthermore, we have not observed significant changes in the analysis due to noise correlations present in real projections. We expect that detector noise could be neglected in similar high exposure cases and become significant in low dose studies. Our MC simulation includes only an X-ray source, object, and detector, and does not contain additional objects that may be in the field of view in industrial applications, such as object holders, conveyor belts, and radiation shielding. These objects could contribute to the scattering signal, especially backscatter, which is not present in our simulation.

We performed defect detection using the MSD architecture for a segmentation DCNN. In the appendix, we show that similar results can be achieved with other DCNN architectures, including DeepLabv3, UNet++, and FPN. Changing metaparameters, such as the number of layers and dilation sequences, may slightly improve the performance for a particular training dataset. Choosing the architecture and fine-tuning the metaparameters are necessary to achieve the best possible performance. Since our goal was to study the effect of scattering, we stopped changing the metaparameters after reaching an accuracy level that could not be easily improved. We have tried to train a classification network instead of segmentation, but the results were inferior. Such a difference in performance can be explained by the absence of the defect mask in the training data, which complicates the learning process. 
We have chosen the MSD architecture due to the relatively small number of weights (compared to other popular architectures such as UNet), the possibility to train an accurate model without a large amount of training data, and a history of successful application to X-ray tasks. Comparison with other DCNN architectures shows that a small number of weights does not significantly reduce accuracy. It is important to note that vision transformers \cite{dosovitskiy2020image} currently outperform DCNNs in standard computer vision tasks (e.g. ImageNet classification), although the novel CNN architectures might compete with transformers \cite{liu2022convnet}.
We used DCNNs due to their successful application to industrial problems and the large number of available implementations. Furthrermore, it is not clear whether vision transformers could outperform DCNNs without large datasets.

When using the difference between POD curves as a metric of performance difference, it is important to note that many POD curves with different feature vectors could be computed for the same network. In preliminary experiments we included object thickness at the defect location and defect distance to the center as parameters in the POD fit. Defect size was chosen as the main POD argument due to a major influence on segmentation accuracy and interpretability. Other variables could be added to Eq. \ref{multivariate_fit}, but the conclusion about the impact of SPR remains the same. Another important assumption is that the chosen training dataset covers all possible combinations of object and defect. Thus, an incorrect segmentation is caused by the input properties and not by encountering a previously unknown defect type or unusual location. Providing sufficient coverage is easier for manufactured objects with known shapes (e.g. castings), and more complicated when both the object and the defect have a large variety in morphology.
The threshold of 50\% $F_1$ score to convert network accuracy into a binary variable is chosen arbitrarily. If increased, it will shift the POD curves to the right with respect to the defect size.

While the presented results are only valid for a selected problem, they illustrate the benefits of using a POD-based methodology. A significant difference in accuracy was only present for a small subset of test projections, and this effect was not visible for other metrics due to variance. Furthermore, POD curves address the concept of the smallest detectable defect, which helps to evaluate whether a system is useful for a particular real-world application. We expect a similar effect of scattering on inspection performance for other problems. If a defect is too small, the contrast in X-ray attenuation is less than the noise level, and the inclusion of scattering does not significantly affect the detection process. If a defect is too large, the contrast would be larger than a variance in the signal due to scattering. Consequently, the biggest influence of the scattering signal should be seen for barely detectable defects when the contrast is similar to the noise level.

It is important to note that any defect detection problem has a number of system properties that affect the noise level and SPR and, consequently, the effect of excluding the scattering signal from the training data. The distance between the object and the detector (known in radiology as the air gap) has a significant effect on SPR values because scattered photons may not reach the detector. Furthermore, a smaller field of view and small objects can reduce the effect of the scattering signal. A longer exposure time decreases the noise level and $s_{90}$, so the effect of scattering as additional noise might become more significant.

\section{Conclusion}

The practical application of data generation techniques for defect detection in industrial products requires a balance between the computational cost of data generation and the resulting accuracy of the algorithm. We have proposed a methodology to quantitatively evaluate the effect of simulating X-ray scattering on data generation for training DCNNs. The POD curves have been used to study the network performance in detail by correlating it with the properties of the test projections. Using a Monte-Carlo simulation algorithm, we have ensured that the difference between the DCNNs was caused only by the presence of the scattering signal. Our performance analysis was successfully applied to the various test cases. The POD curves have shown under which test conditions the difference between DCNNs is significant and when it is negligible. In particular, we have shown how the scattering-to-primary ratio affects the network accuracy and the influence of the data generation approach. The presented methodology can be used to decide if the defect detection performance is sufficient for a particular task and what level of accuracy in data generation is necessary to achieve that performance.

\section*{Code and Data availability}
The code for Monte-Carlo simulations can be found on Github: \url{https://github.com/vandriiashen/mc-scattering}. The training and testing data were generated using this code. Sequences of parameters used for volume generation are included in the repository. Scripts for DCNN training and POD computation are available at \url{https://github.com/vandriiashen/pod-xray-images}.

\section*{Appendix: Different DCNN architectures}

\begin{figure}[p]
\centering
\includegraphics[width=1.\linewidth]{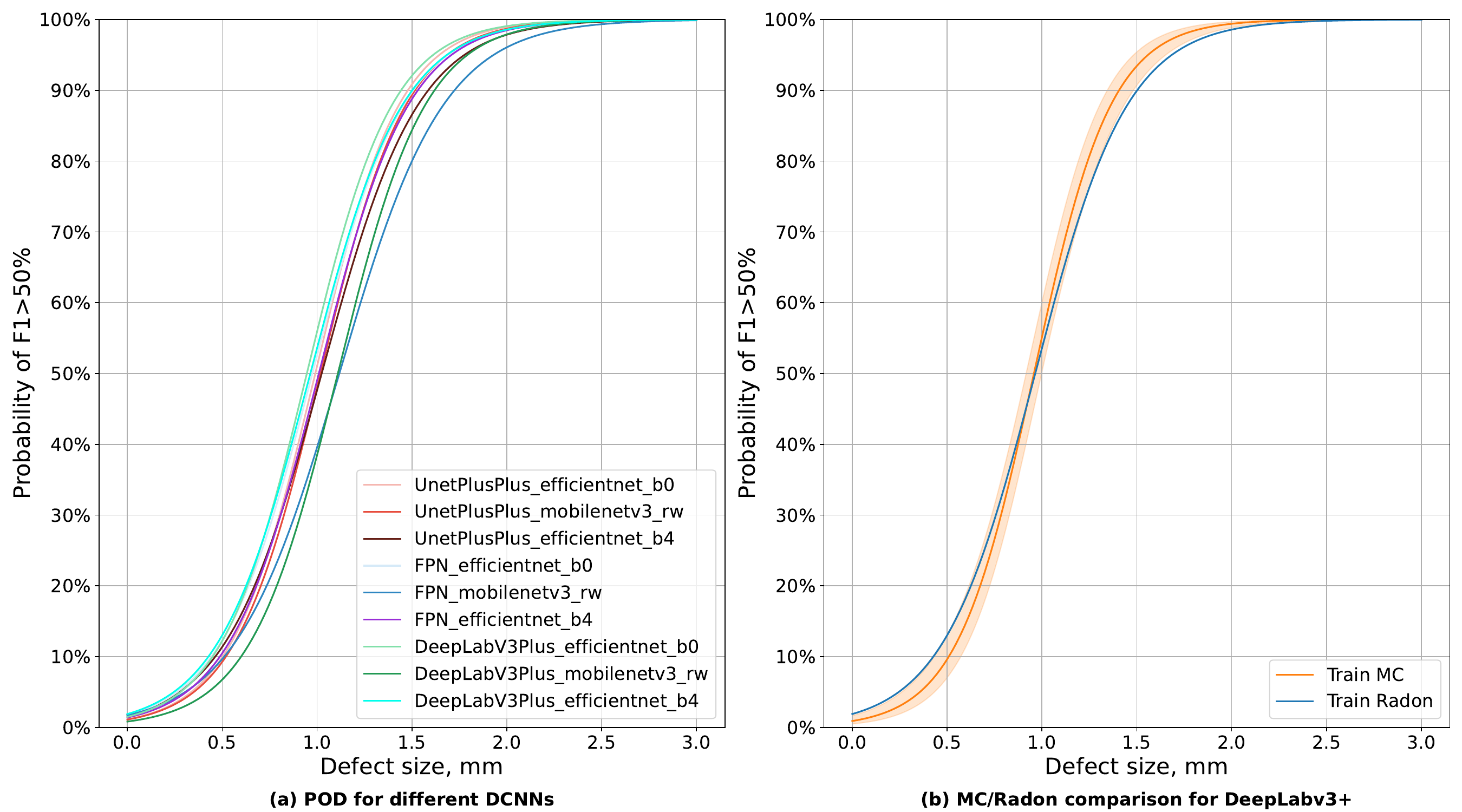}
\caption{Comparison of POD curves for the DCNNs with different architectures tested on the iron dataset at 450~kV. DeepLabv3+ outperforms other architectures tested, MSD achieves one of the highest accuracy levels despite having the least number of parameters. The difference in $s_{90}$ is around 10\% when comparing the best and the worst model (a). Comparison of the networks trained on MC and Radon data when the network uses the DeepLabv3+ architecture with the EfficientNetB0 encoder. The decrease in $s_{90}$ is similar to MSD (b). }
\label{comp_dcnn}
\end{figure}

For the dataset of iron at 450~kV we have trained DCNNs with different state-of-the-art architectures using the Segmentation Models package \cite{Iakubovskii:2019}. We have tried UNet++, FPN (Feature Pyramid Network), and DeepLabv3+ as semantic segmentation architectures, and EfficientNet, MobileNet, and ResNet (not included in the results because the performance was significantly worse) as encoders. The accuracy of each network on the test dataset was evaluated with POD curves in the same way as for MSD (Fig. \ref{comp_dcnn}a). The networks have similar performance, the difference in $s_{90}$ between the best and the worst model is around 10\%. The difference between training on Radon and MC data can be evaluated for other network architectures, and the effect is similar to MSD. For example, Fig. \ref{comp_dcnn}b is made with the same data as Fig. \ref{comp_mat}c, but the network architecture is DeepLabv3+ instead of MSD. 

The similarity in network performance indicates that the difficulty of segmentation in the model problem stems from the image and noise properties and not from deep learning. This supports the assumption that the generated projections have enough variety to train a DCNN.

%\begin{figure}[ht]

\bibliographystyle{unsrt} 
\bibliography{main}

\begin{thebibliography}{10}

\bibitem{ISO_xray}
{Non-destructive testing — Industrial computed radiography with storage phosphor imaging plates — Part 2: General principles for testing of metallic materials using X-rays and gamma rays}.
\newblock Standard, International Organization for Standardization, Geneva, CH, 09 2017.

\bibitem{chen2021interfacial}
Hongbing Chen, Xin Nie, Shiyu Gan, Yudong Zhao, and Huahua Qiu.
\newblock Interfacial imperfection detection for steel-concrete composite structures using ndt techniques: A state-of-the-art review.
\newblock {\em Engineering Structures}, 245:112778, 2021.

\bibitem{mery2015computer}
Domingo Mery.
\newblock Computer vision for {X-R}ay testing.
\newblock {\em Switzerland: Springer International Publishing}, 10:978--3, 2015.

\bibitem{yang2020using}
Jing Yang, Shaobo Li, Zheng Wang, Hao Dong, Jun Wang, and Shihao Tang.
\newblock Using deep learning to detect defects in manufacturing: a comprehensive survey and current challenges.
\newblock {\em Materials}, 13(24):5755, 2020.

\bibitem{lee2021review}
Kidong Lee, Sung Yi, Soongkeun Hyun, and Cheolhee Kim.
\newblock Review on the recent welding research with application of {CNN}-based deep learning part {I}: Models and applications.
\newblock {\em Journal of Welding and Joining}, 39(1):10--19, 2021.

\bibitem{mery2015gdxray}
Domingo Mery, Vladimir Riffo, Uwe Zscherpel, German Mondrag{\'o}n, Iv{\'a}n Lillo, Irene Zuccar, Hans Lobel, and Miguel Carrasco.
\newblock {GDXray}: The database of {X}-ray images for nondestructive testing.
\newblock {\em Journal of Nondestructive Evaluation}, 34:1--12, 2015.

\bibitem{bellon2007artist}
Carsten Bellon and Gerd-R{\"u}diger Jaenisch.
\newblock a{RT}ist--analytical {RT} inspection simulation tool.
\newblock In {\em Proc DIR}, pages 25--27, 2007.

\bibitem{gong2018rapid}
Qian Gong, Razvan-Ionut Stoian, David~S Coccarelli, Joel~A Greenberg, Esteban Vera, and Michael~E Gehm.
\newblock Rapid simulation of {X}-ray transmission imaging for baggage inspection via {GPU}-based ray-tracing.
\newblock {\em Nuclear Instruments and Methods in Physics Research Section B: Beam Interactions with Materials and Atoms}, 415:100--109, 2018.

\bibitem{gong2019rapid}
Qian Gong, Joel~A Greenberg, Razvan-Ionut Stoian, David Coccarelli, Esteban Vera, and Michael~E Gehm.
\newblock Rapid simulation of {X}-ray scatter measurements for threat detection via {GPU}-based ray-tracing.
\newblock {\em Nuclear Instruments and Methods in Physics Research Section B: Beam Interactions with Materials and Atoms}, 449:86--93, 2019.

\bibitem{georgiou2007pod}
George~A Georgiou.
\newblock {PoD} curves, their derivation, applications and limitations.
\newblock {\em Insight-Non-Destructive Testing and Condition Monitoring}, 49(7):409--414, 2007.

\bibitem{tyystjarvi2022automated}
Topias Tyystj{\"a}rvi, Iikka Virkkunen, Peter Fridolf, Anders Rosell, and Zuheir Barsoum.
\newblock Automated defect detection in digital radiography of aerospace welds using deep learning.
\newblock {\em Welding in the World}, 66(4):643--671, 2022.

\bibitem{yosifov2022probability}
M~Yosifov, M~Reiter, S~Heupl, C~Gusenbauer, B~Fr{\"o}hler, R~Fern{\'a}ndez-Guti{\'e}rrez, J~De~Beenhouwer, J~Sijbers, J~Kastner, and C~Heinzl.
\newblock Probability of detection applied to x-ray inspection using numerical simulations.
\newblock {\em Nondestructive Testing and Evaluation}, 37(5):536--551, 2022.

\bibitem{masci2012steel}
Jonathan Masci, Ueli Meier, Dan Ciresan, J{\"u}rgen Schmidhuber, and Gabriel Fricout.
\newblock Steel defect classification with max-pooling convolutional neural networks.
\newblock In {\em The 2012 international joint conference on neural networks (IJCNN)}, pages 1--6. IEEE, 2012.

\bibitem{aydin2018new}
Ilhan Aydin, Mehmet Karakose, and AKIN Erhan.
\newblock A new approach for baggage inspection by using deep convolutional neural networks.
\newblock In {\em 2018 International Conference on Artificial Intelligence and Data Processing (IDAP)}, pages 1--6. IEEE, 2018.

\bibitem{parlak2023deep}
{\.I}smail~Enes Parlak and Erdal Emel.
\newblock Deep learning-based detection of aluminum casting defects and their types.
\newblock {\em Engineering Applications of Artificial Intelligence}, 118:105636, 2023.

\bibitem{bian2016multiscale}
Xiao Bian, Ser~Nam Lim, and Ning Zhou.
\newblock Multiscale fully convolutional network with application to industrial inspection.
\newblock In {\em 2016 IEEE winter conference on applications of computer vision (WACV)}, pages 1--8. IEEE, 2016.

\bibitem{van2022inline}
Tim Van De~Looverbosch, Jiaqi He, Astrid Tempelaere, Klaas Kelchtermans, Pieter Verboven, Tinne Tuytelaars, Jan Sijbers, and Bart Nicolai.
\newblock Inline nondestructive internal disorder detection in pear fruit using explainable deep anomaly detection on {X}-ray images.
\newblock {\em Computers and Electronics in Agriculture}, 197:106962, 2022.

\bibitem{tempelaere2023synthetic}
Astrid Tempelaere, Tim Van De~Looverbosch, Klaas Kelchtermans, Pieter Verboven, Tinne Tuytelaars, and Bart Nicolai.
\newblock Synthetic data for {X}-ray {CT} of healthy and disordered pear fruit using deep learning.
\newblock {\em Postharvest Biology and Technology}, 200:112342, 2023.

\bibitem{armanious2020medgan}
Karim Armanious, Chenming Jiang, Marc Fischer, Thomas K{\"u}stner, Tobias Hepp, Konstantin Nikolaou, Sergios Gatidis, and Bin Yang.
\newblock Med{GAN}: Medical image translation using {GAN}s.
\newblock {\em Computerized medical imaging and graphics}, 79:101684, 2020.

\bibitem{agostinelli2003geant4}
Sea Agostinelli, John Allison, K~al Amako, John Apostolakis, H~Araujo, Pedro Arce, Makoto Asai, D~Axen, Swagato Banerjee, GJNI Barrand, et~al.
\newblock {GEANT4} — a simulation toolkit.
\newblock {\em Nuclear instruments and methods in physics research section A: Accelerators, Spectrometers, Detectors and Associated Equipment}, 506(3):250--303, 2003.

\bibitem{jan2011gate}
S{\'e}bastien Jan, Didier Benoit, E~Becheva, Thomas Carlier, Franca Cassol, Patrice Descourt, T~Frisson, L~Grevillot, L~Guigues, L~Maigne, et~al.
\newblock {GATE} v6: a major enhancement of the gate simulation platform enabling modelling of ct and radiotherapy.
\newblock {\em Physics in Medicine \& Biology}, 56(4):881, 2011.

\bibitem{sun2010improved}
Miguel Sun and JM~Star-Lack.
\newblock Improved scatter correction using adaptive scatter kernel superposition.
\newblock {\em Physics in Medicine \& Biology}, 55(22):6695, 2010.

\bibitem{bhatia2016scattering}
Navnina Bhatia, David Tisseur, Fanny Buyens, and Jean~Michel L{\'e}tang.
\newblock Scattering correction using continuously thickness-adapted kernels.
\newblock {\em NDT \& E International}, 78:52--60, 2016.

\bibitem{maier2018deep}
Joscha Maier, Stefan Sawall, Michael Knaup, and Marc Kachelrie{\ss}.
\newblock Deep scatter estimation ({DSE}): Accurate real-time scatter estimation for {X}-ray {CT} using a deep convolutional neural network.
\newblock {\em Journal of Nondestructive Evaluation}, 37:1--9, 2018.

\bibitem{ruhrnschopf2011general}
Ernst-Peter R{\"u}hrnschopf and Klaus Klingenbeck.
\newblock A general framework and review of scatter correction methods in x-ray cone-beam computerized tomography. part 1: scatter compensation approaches.
\newblock {\em Medical physics}, 38(7):4296--4311, 2011.

\bibitem{barnes1991contrast}
Gary~T Barnes.
\newblock Contrast and scatter in x-ray imaging.
\newblock {\em Radiographics}, 11(2):307--323, 1991.

\bibitem{cardoso2009evaluation}
Simone Cardoso, Odair Gon{\c{c}}alves, and Helio Schechter.
\newblock Evaluation of scatter-to-primary ratio in radiological conditions.
\newblock {\em Applied Radiation and Isotopes}, 67(4):544--548, 2009.

\bibitem{cunha2010evaluation}
DM~Cunha, A~Tomal, and Mart{\'\i}n~Eduardo Poletti.
\newblock Evaluation of scatter-to-primary ratio, grid performance and normalized average glandular dose in mammography by {Monte Carlo} simulation including interference and energy broadening effects.
\newblock {\em Physics in Medicine \& Biology}, 55(15):4335, 2010.

\bibitem{boone1988analytical}
John~M Boone and J~Anthony Seibert.
\newblock An analytical model of the scattered radiation distribution in diagnostic radiology.
\newblock {\em Medical physics}, 15(5):721--725, 1988.

\bibitem{martin1999measurement}
CJ~Martin, PF~Sharp, and DG~Sutton.
\newblock Measurement of image quality in diagnostic radiology.
\newblock {\em Applied radiation and isotopes}, 50(1):21--38, 1999.

\bibitem{jessen2004balancing}
Karl~Arne Jessen.
\newblock Balancing image quality and dose in diagnostic radiology.
\newblock {\em European Radiology Supplements}, 14:9--18, 2004.

\bibitem{andriiashen2023ct}
Vladyslav Andriiashen, Robert van Liere, Tristan van Leeuwen, and K~Joost Batenburg.
\newblock {CT}-based data generation for foreign object detection on a single x-ray projection.
\newblock {\em Scientific Reports}, 13(1):1881, 2023.

\bibitem{hendrycks2017a}
Dan Hendrycks and Kevin Gimpel.
\newblock A baseline for detecting misclassified and out-of-distribution examples in neural networks.
\newblock International Conference on Learning Representations, ICLR, 2017.

\bibitem{hendrycks2021many}
Dan Hendrycks, Steven Basart, Norman Mu, Saurav Kadavath, Frank Wang, Evan Dorundo, Rahul Desai, Tyler Zhu, Samyak Parajuli, Mike Guo, et~al.
\newblock The many faces of robustness: A critical analysis of out-of-distribution generalization.
\newblock In {\em Proceedings of the IEEE/CVF International Conference on Computer Vision}, pages 8340--8349, 2021.

\bibitem{whiting2002signal}
Bruce~R Whiting.
\newblock Signal statistics in x-ray computed tomography.
\newblock In {\em Medical Imaging 2002: Physics of Medical Imaging}, volume 4682, pages 53--60. SPIE, 2002.

\bibitem{freud2005hybrid}
N~Freud, J-M L{\'e}tang, and D~Babot.
\newblock A hybrid approach to simulate multiple photon scattering in x-ray imaging.
\newblock {\em Nuclear Instruments and Methods in Physics Research Section B: Beam Interactions with Materials and Atoms}, 227(4):551--558, 2005.

\bibitem{hernandez2016xpecgen}
Guillermo Hern{\'a}ndez and Francisco Fern{\'a}ndez.
\newblock xpecgen: A program to calculate x-ray spectra generated in tungsten anodes.
\newblock {\em J. Open Source Softw.}, 1(7):62, 2016.

\bibitem{whiting2006properties}
Bruce~R Whiting, Parinaz Massoumzadeh, Orville~A Earl, Joseph~A O'Sullivan, Donald~L Snyder, and Jeffrey~F Williamson.
\newblock Properties of preprocessed sinogram data in x-ray computed tomography.
\newblock {\em Medical physics}, 33(9):3290--3303, 2006.

\bibitem{smalley2020image}
Duane Smalley, Stuart Baker, Brandon Baldonado, Jesus Castaneda, Andrew Corredor, Jessica~H Clayton, Logan Fegenbush, Cort Gautier, Amanda Gehring, Todd Haines, et~al.
\newblock Image restoration of high-energy x-ray radiography with a scintillator blur model.
\newblock {\em Nuclear Instruments and Methods in Physics Research Section A: Accelerators, Spectrometers, Detectors and Associated Equipment}, 968:163910, 2020.

\bibitem{pelt2018mixed}
Dani{\"e}l~M Pelt and James~A Sethian.
\newblock A mixed-scale dense convolutional neural network for image analysis.
\newblock {\em Proceedings of the National Academy of Sciences}, 115(2):254--259, 2018.

\bibitem{hendriksen-2019-msd-pytor}
Allard~A. Hendriksen.
\newblock ahendriksenh/msd\_pytorch: v0.7.2, December 2019.

\bibitem{seabold2010statsmodels}
Skipper Seabold and Josef Perktold.
\newblock statsmodels: Econometric and statistical modeling with python.
\newblock In {\em 9th Python in Science Conference}, 2010.

\bibitem{chen2021multivariate}
Guanren Chen, Yijun Guo, Takuya Katagiri, Haicheng Song, Takuma Tomizawa, Noritaka Yusa, and Hidetoshi Hashizume.
\newblock Multivariate probability of detection (pod) analysis considering the defect location for long-range, non-destructive pipe inspection using electromagnetic guided wave testing.
\newblock {\em NDT \& E International}, 124:102539, 2021.

\bibitem{sarno2018normalized}
A~Sarno, G~Mettivier, F~Di~Lillo, RM~Tucciariello, K~Bliznakova, and P~Russo.
\newblock Normalized glandular dose coefficients in mammography, digital breast tomosynthesis and dedicated breast ct.
\newblock {\em Physica Medica}, 55:142--148, 2018.

\bibitem{di2020geant4}
Francesco di~Franco, A~Sarno, G~Mettivier, AM~Hernandez, K~Bliznakova, JM~Boone, and P~Russo.
\newblock Geant4 monte carlo simulations for virtual clinical trials in breast x-ray imaging: Proof of concept.
\newblock {\em Physica Medica}, 74:133--142, 2020.

\bibitem{coban2020explorative}
Sophia~Bethany Coban, Felix Lucka, Willem~Jan Palenstijn, Denis Van~Loo, and Kees~Joost Batenburg.
\newblock Explorative imaging and its implementation at the flex-ray laboratory.
\newblock {\em Journal of Imaging}, 6(4):18, 2020.

\bibitem{dosovitskiy2020image}
Alexey Dosovitskiy, Lucas Beyer, Alexander Kolesnikov, Dirk Weissenborn, Xiaohua Zhai, Thomas Unterthiner, Mostafa Dehghani, Matthias Minderer, Georg Heigold, Sylvain Gelly, et~al.
\newblock An image is worth 16x16 words: Transformers for image recognition at scale.
\newblock {\em arXiv preprint arXiv:2010.11929}, 2020.

\bibitem{liu2022convnet}
Zhuang Liu, Hanzi Mao, Chao-Yuan Wu, Christoph Feichtenhofer, Trevor Darrell, and Saining Xie.
\newblock A convnet for the 2020s.
\newblock In {\em Proceedings of the IEEE/CVF conference on computer vision and pattern recognition}, pages 11976--11986, 2022.

\bibitem{Iakubovskii:2019}
Pavel Iakubovskii.
\newblock Segmentation {M}odels {P}ytorch.
\newblock \url{https://github.com/qubvel/segmentation_models.pytorch}, 2019.

\end{thebibliography}

\end{document}